\documentclass[useAMS,usenatbib]{mn2e} 
\usepackage{graphicx}

\newcommand{\aap}{A\&A}  
  
\newcommand{\aj}{AJ}  
\newcommand{\apj}{ApJ}

\newcommand{\mnras}{MNRAS}

\def \kms{\ifmmode{~{\rm km\,s}^{-1}}\else{~km~s$^{-1}$}\fi}  
\def \vhel{\ifmmode{V_{{\rm hel}}}\else{$V_{{\rm hel}}$}\fi}  
\def \vsys{\ifmmode{V_{{\rm sys}}}\else{$V_{{\rm sys}}$}\fi}  
\def \vlsr{\ifmmode{V_{{\rm lsr}}}\else{$V_{{\rm lsr}}$}\fi}  
\def \vobs{\ifmmode{V_{{\rm obs}}}\else{$V_{{\rm obs}}$}\fi}  
\def \degree{\ifmmode{^{\circ}}\else{$^{\circ}$}\fi}  
\def \lsun{\ifmmode{{\rm\ L}_\odot}\else{${\rm\ L}_\odot $}\fi}  
\def \msun{\ifmmode{{\rm\ M}_\odot}\else{${\rm\ M}_\odot$}\fi}  
\def \myr{\ifmmode{{\rm\ M}_\odot{\rm\ yr}^{-1}}\else{${\rm\ M}_\odot$   
yr$^{-1}$}\fi}  
\def \teff{\ifmmode{{\rm{T}}_{\rm eff}}\else{${\rm{T}}_{\rm eff}$}\fi}  
\def \mdot{\ifmmode{{\rm\dot{M}}}\else{${\rm\dot{M}}$}\fi}

\newcommand{\ha}{H$\alpha$}

\newcommand{\nii}{[N~{\sc ii}]~6584~\AA}

\newcommand{\NII}{[N~{\sc ii}]~6548~\&~6584~\AA}  
  
\newcommand{\oiii}{[O~{\sc iii}]~5007~\AA}

\title[Cometary tails in Helix]{Flows along cometary tails 
in the Helix planetary nebula NGC~7293}  
\author[J. Meaburn and P. Boumis]{J. Meaburn$^{1}$\thanks{E-mail:
jmeaburn@jb.man.ac.uk} and P. Boumis$^{2}$\\
$^{1}$Jodrell Bank Centre for Astrophysics, University of Manchester,  
Manchester M13 9PL, UK.\\  
$^{2}$Institute of Astronomy \& Astrophysics, National Observatory of  
Athens, I. Metaxa \& V. Paulou, GR--152 36 P. Penteli, Athens,  
Greece.\\  
}  
  
\begin{document}  
  
\date{Accepted 2009 October 15. Received 2009 September 23; in original
form 2009 August 11}
  
\pagerange{\pageref{firstpage}--\pageref{lastpage}} \pubyear{2009}  
  
\maketitle  
  
\label{firstpage}  
  
\begin{abstract}  
Previous velocity images which reveal flows of ionized gas along the
most prominent cometary tail (from Knot 38) in the Helix planetary
nebula are compared with that taken at optical wavelengths with the
Hubble Space Telescope and with an image in the emission from
molecular hydrogen. The flows from the second most prominent tail from
Knot 14 are also considered. The kinematics of the tail from the more
complex Knot 32, shown here for the first time, also reveals an
acceleration away from the central star.  All of the tails are
explained as accelerating ionized flows of ablated material driven by
the previous, mildly supersonic, AGB wind from the central star. The
longest tail of ionized gas, even though formed by this mechanism in a
very clumpy medium, as revealed by the emission from molecular
hydrogen, appears to be a coherent outflowing feature.
\end{abstract}  
  
  
\section{Introduction}  

Dense, neutral (\citealt{mea92}; \citealt{hug92}) knots with ionized
cometary tails are found in the central regions of planetary nebulae
(PNe); the most famous being those in the Helix nebula for, at a
distance of only 213 $+$30/$-$16 pc \citep{har07}, they were easily
detected and resolved by early ground--based observations (Baade and
reported by \citealt{vor68}). The incredibly clumpy nature of the
neutral material in the disk of the Helix nebula (when modelled as a
bi--polar PNe viewed along its axis, \citealt{mea98}; \citealt{mea05})
has now been revealed in spectacular fashion in the imagery in the
H$_{2}$ emission line by \citet{mat09}. Previously, \citet{mei05} had
estimated that there were 23,000 cometary knots and that inevitably
tail interactions must be occurring. \citet{mat09} show conclusively
that it is an inner region in the Helix disk, towards the central
star, where tails are observed from neutral globules surrounded by an
outer clumpy region free of such tails.

\citet{dys03} reviewed the two broad models for the creation of
the cometary tails; either they are shadowed from the ionizing
radiation of the central star, and their surfaces photo-ionized by
scattered Lyman photons in the nebula, or they are dynamically
produced as the particle winds from the central star swept past the
slowly expanding system of dense globules.

A critical distinction between these models arises if the cometary
tails can be seen to be flowing along their lengths away from their
parent globules. Only the `dynamic' model would have this effect. In
fact, a numerical simulation of the flow of a moderately supersonic
particle wind past the ionized head of a globule \citep{dys06}
predicted not only the creation of a cometary tail but a moderate
acceleration of ionized material parallel to the tail surface and away
from the globule. In this initial model the neutral material away from
the globule had a smooth density distribution.  

The most comprehensive optical study of the kinematics of the Helix
system of globules remains that described in \citet{mea98}. Here,
spatially resolved \nii\ profiles, from 300 separate long-slit
(163\arcsec\ long) positions, were obtained with the Manchester
echelle spectrometer (MES -- \citealt{mea84} but now with a CCD as the
detector) on the Anglo--Australian telescope in exceptional `seeing'
conditions, over three separate `blocks' of globules and their tails
in the nebular core. Furthermore, a kinematical `case study' of the
globule which is apparently the 2nd closest to the nebular core (Knot
14), and a length of its tail, was carried out in a range of emission
lines. The principal outcome was to show that the system of central
globules is concentrated in a disk expanding itself at 14\kms. Flows
parallel to the surfaces of the cometary tails of two of the most
prominent globules (Knots 38 and 14) were also detected and even an
acceleration of this flow along the length of the longest tail (from
Knot 38) suggested. \citet{ode07} challenged these latter assertions
without making any further kinematical observations but simply because
they claim that more recent HST images reveal confusing minor globules
in the longest tail (Knot 38). With the dismissal of these kinematical
effects they proceeded to support the shadowing theories of the
creation of the tails.
The aim of the present paper is to reassess the strength of the
deductions from the \citet{mea98} kinematical data set in the light of
the subsequent HST and very latest H$_{2}$ imagery by \citet{mat09}. 
In particular, to
consider if the evidence in \citet{mea98} for flows along the tail
surfaces of Knots 38 and 14, away from the central star, is now
invalidated by this HST imagery alone as suggested by
\citet{ode07}. Furthermore, the flow behind Knot 32, which was not
considered hitherto, is now presented to strengthen the original
suggestion that accelerating flows in the cometary tails could be
ubiquitous though in a clumpy medium.
\section[]{Knots 38 and 14}  
Knot 38 has the longest (62\arcsec) cometary tail and is the
apparently closest knot to the central star of NGC 7293, while Knot 14
is the next closest. The heads of both have arcs of \oiii\ emission
facing this star which indicates that they protrude into the hard
radiation field of the central volume of the bi--polar,
ellipsoidally--shaped, nebula that produces the characteristic helical
appearance in emission lines of lower ionization species
(\citealt{mea98}; \citealt{mea05}). The HST archival images (PI NAME:
Meixner, PID:9700) of Knots 38 (HST ACS/WFC J8KR14040) and 14 (HST
ACS/WFC J8KR0840) are presented respectively in Figs.  1a and 2.  These
should be compared with those taken with the New Technology Telescope
(NNT - Chile) and presented in Meaburn et al (1998). All are in the
light of the \ha\ plus \NII\ nebular emission lines.

The uniquely long ionized tail of Knot 38 in Fig. 1 is prominent
in both the HST and former \citep{mea98} NTT images and appears to be
a coherent structure in both i.e. it is not simply a consequence of
chance superposition of a large number of fore-- or background tails
along the same sightlines. Minor ionized knots appear along the tail
in the image in Fig. 1a but it is the H$_{2}$ 2.12$\mu$m 
image \citep{mat09} shown in Fig. 1b that
emphasises that the tail of ionized gas, being so long, is formed
around an internal core composed of a large number of minor clumps of
neutral material.

With this as a starting point the kinematics along this tail should be
re--considered by examining the \nii\ line profiles presented in
\citet{mea98}. The centroids of these profiles are shown in fig. 13 of
that paper to change along the 62\arcsec\ length of this tail by a
radial velocity difference (from central knot to tail end) by
10\kms. The velocity images in figs. 4 and 12 of the same paper
confirm this systematic radial velocity change in a different
way. Unfortunately, the four images in different heliocentric radial
velocity (\vhel) ranges became jumbled in the production of fig. 12 in
the \citet{mea98} paper. These should be \vhel\ $= -$31 to $-$27 (top
right), $-$25 to $-$21 (top left), $-$20 to $-$15 (bottom right) and
$-$14 to $-$10 (bottom left) and all \kms. When these four images are
considered with this correction it is clear that the head of Knot 38
appears alone in the top right frame then progressively the tail
appears in the subsequent three frames towards more positive
velocities as the image of the head declines. The acceleration along
the tail length, if regarded as a coherent feature starting at Knot
38, is very clear and not dominated by the minor confusing knots
apparent in Fig. 1a. Those areas free of these along the tail length of
Knot 38 in Fig. 1 clearly show this radial velocity change. The appearance
of the neutral material in the tail of Knot 38 and seen in Fig. 1b
suggests that the ionized outflow is in a sheath around a clumpy neutral
core.

Similarly, the \nii\ line profiles up to 5\arcsec\ from the head of
Knot 14 (fig. 9 of \citealt{mea98}) show a systematic change of radial
velocity of their centroids of -6 \kms. This was not covered by the H$_{2}$
imagery of \citep{mat09}. 
The tail from Knot 14  was modelled
kinematically as a flow parallel to the globule and tail
surfaces. The HST  images in Fig. 2 show that there are no significant
minor ionized knots along this small length of the tail.
 Again an
accelerating flow away from the central star is indicated. If tilted
at 25\degree\ to the plane of the sky this change of outflow velocity
amounts to 14 \kms\ compared to the apex of the head of the cometary
knot.
\section{Knot 32} 
As shown in Figs. 1a \& b
  Knot 32 has a more complex structure than that of
both Knots 38 and 14. Again, a faint tail of diffuse ionized gas
can be seen extending radially away from the central star though the 
central knot is composed of several neutral clumps. This tail of course
could be an unrelated feature along the same sightline but its appearance
suggests that this is unlikely. The
kinematics of the Knot 32 tail have now been examined using the set of
profiles of the \nii\ emission line from 100 longslit positions. These
have an EW orientation and each is separated from its neighbour by
1\arcsec. The slit width is $\equiv$ 6\kms\ and 0.5\arcsec\ on the
sky. These observations and their analysis are described fully in
\citet{mea98} and the reader is referred there for detailed
information.

However, the \nii\ line profile from the very apex of this knot,
facing the ionizing star, is shown in Fig. 3a and that of the knot
head (1.5 \arcsec\ further from this star) in Fig. 3b. The \nii\
profiles from the faint material of the knot tail ( 14\arcsec\ and
16\arcsec\ away from the apex) are shown in Figs. 3c \& d
respectively. The positions where these line profiles were
obtained are marked a--d in the velocity  image in Fig. 4c.
The \nii\ profiles from the host nebula have been subtracted in each
case leaving only the profiles of the \nii\ emission from Knot 32 and
its tail in those shown in Fig. 3a--d.  This type of `velocity
imagery' is inevitably blurred in the NS dimension for the 
adjacent longslit spectra, each orientated EW, that are used to generate
individual velocity images are separated by 1\arcsec\ which is then
convolved in the imagery with the angular slit width and the
0.8\arcsec\ seeing disk.  
A distinct positive shift of around 8 \kms\
in radial velocity can be seen to be occurring between the apex of the
knot and the material in the extended tail.

The `velocity' images from the same data set in Figs. 4a--c show this
radial velocity shift in a different way. The head of the knot alone
appears in Fig. 4a (\vhel\ $= -$31 to $-$27\kms), the nearest part of
the tail in Fig. 4b (\vhel\ $= -$24 to $-$21\kms) and the furthest
extent of this tail in Fig. 4c (\vhel\ $= -$20 to $-$16\kms) all of
which is consistent with the profiles shown in Fig. 3a--d.

\section[]{Conclusions}
The 62\arcsec\ long tail from Knot 38 appears to be a coherent
structure in the HST, NTT and the previous 
velocity imagery but formed around an
extremely clumpy neutral medium as shown by the H$_{2}$ imagery.

The evidence that there are accelerating flows along the walls of the
tails of the two most prominent cometary globules in NGC 7293 (Knots
38 and 14), away from the central star, is not substantially affected
by the higher resolution HST images or even the extremely clumpy
nature of the neutral material revealed in the H$_{2}$ imagery.

A similar flow away from the more complex Knot 32 is shown 
to be occurring for the first time in the present paper.

It is valid to attempt to explain flows along the cometary tails in
the dynamical type of model in \citet{dys06} but now
modified to accommodate a very clumpy ambient medium similar
to that in \citet{pit05} which had been derived for 
more general clumpy phenomena.

It is becoming very clear \citep{mat09} that the AGB particle wind
overflowing the clumpy neutral structure of the disk of NGC 7293, at
mildly supersonic velocity, can create and accelerate the cometary
tails. Previously, \citet{mea82} had shown that an inner
quasi--spherical volume of material shielded the system of cometary
knots from exposure to any subsequent fast stellar particle wind.
  
\section*{Acknowledgments}  
  
We would like to thank the referee for constructive comments that have
improved the paper considerably. We also thank M. Matsuura who kindly
provided  their  molecular hydrogen image of Helix in FITS format. The
observations made with the NASA/ESA Hubble Space Telescope, obtained
from the data archive at the Space Telescope Institute. STScI is
operated by the association of Universities for Research in Astronomy,
Inc. under the NASA contract NAS 5-26555.

%

 
\newpage  
  
\begin{figure*}
\centering
\scalebox{0.75}{\includegraphics{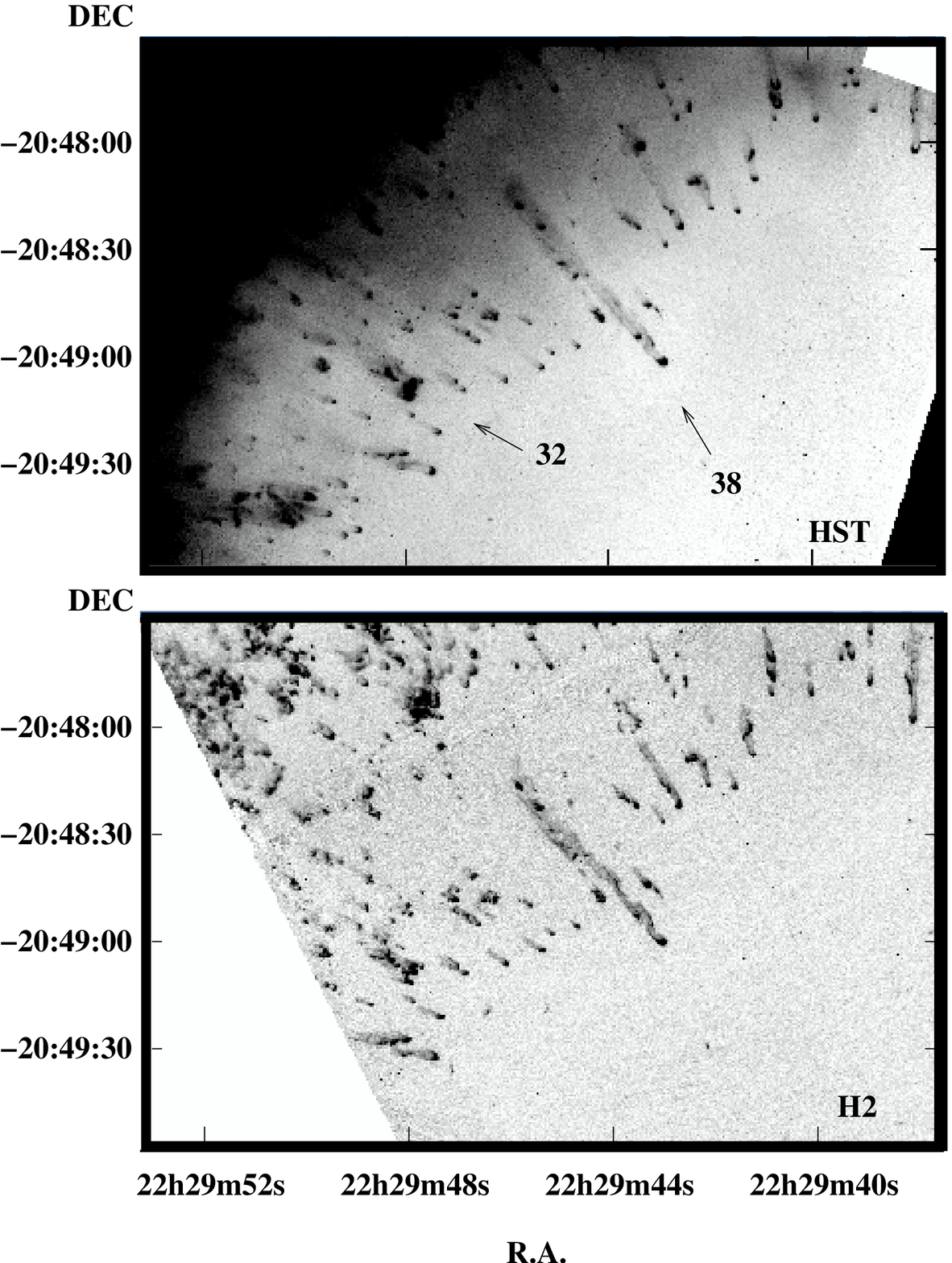}}  
\caption[]{a) Image in the light of the \ha\ and \NII\
nebular emission lines of the region of Knot 38
with the Hubble Space Telescope. The central star is at RA (22h 29m 38.55s)
DEC (-20\degree\ 50\arcmin\ 13.6\arcsec) (J2000).b) Image in the light
of the 2.12 $\mu$m H$_{2}$  emission line  \citep{mat09}.}
\label{fig1} 
\end{figure*}
\begin{figure*}
\scalebox{0.75}{\includegraphics{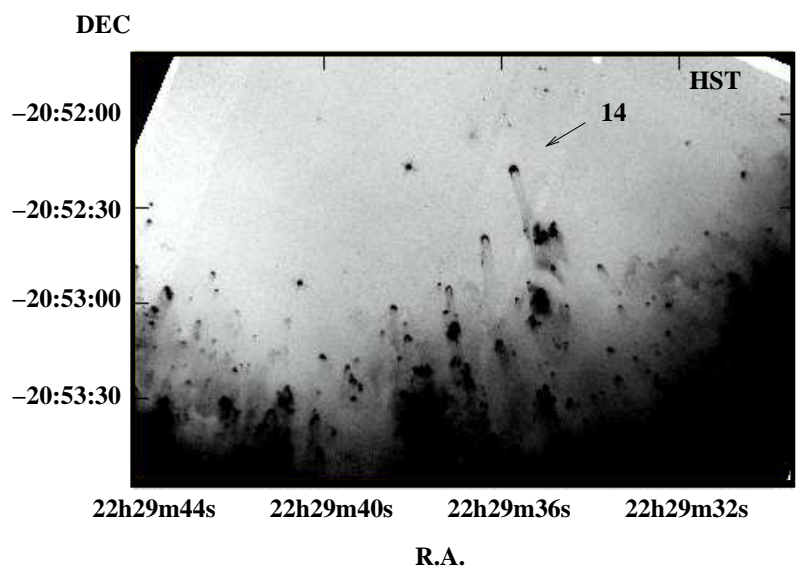}}  
\caption{As for Fig. 1a but of the region of Knot 14}
\label{fig2} 
\end{figure*}

\begin{figure*}
\scalebox{0.75}{\includegraphics{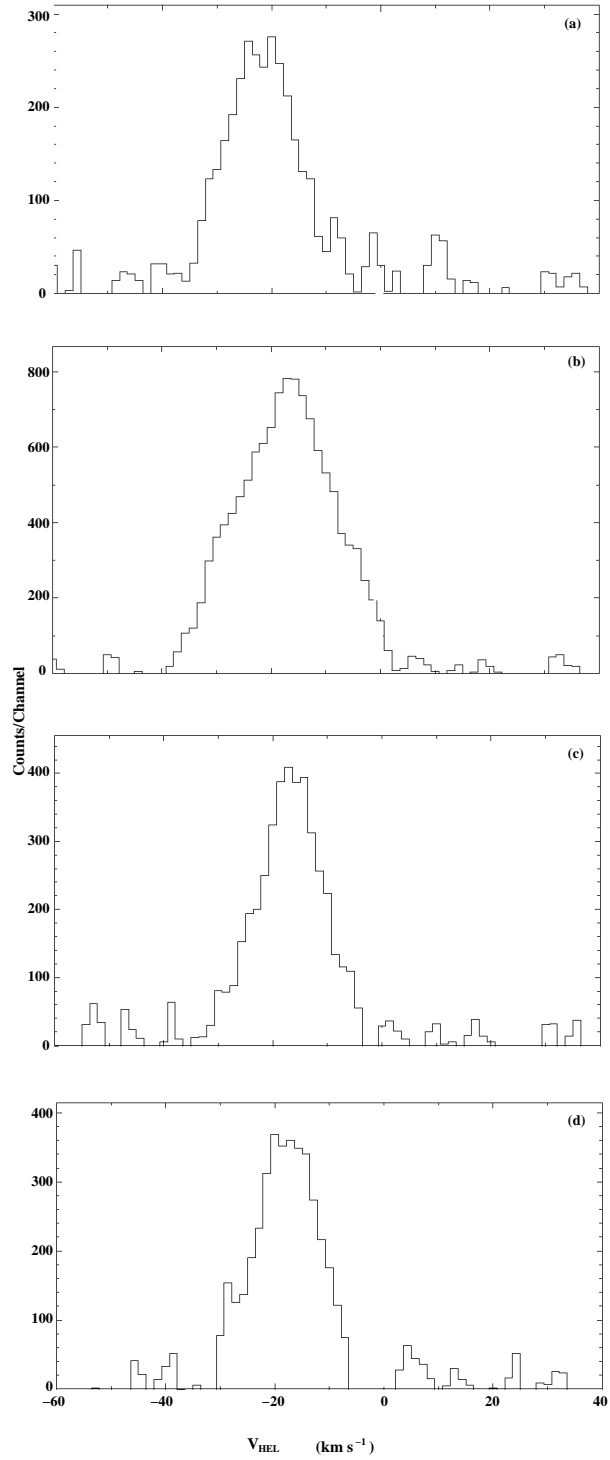}}  
\caption{\nii\ line profiles are shown. That in a) is from the apex of
Knot 32 closest to the central star, and respectively in b), c) and d)
1.5\arcsec, 14\arcsec\ and 16\arcsec\ radially away from this position
i.e. towards the knot tail. The profiles in a--d are from EW
rectangular windows 3.8\arcsec, 3.8\arcsec, 5.4\arcsec\ and
8.9\arcsec\ long respectively and each 0.5\arcsec\ wide. The line
profiles from adjacent slit lengths have been subtracted for each
profile shown here to eliminate the confusing profiles of the general
nebulosity along the same sightlines.}
\label{fig3} 
\end{figure*}
\begin{figure*}
\scalebox{0.75}{\includegraphics{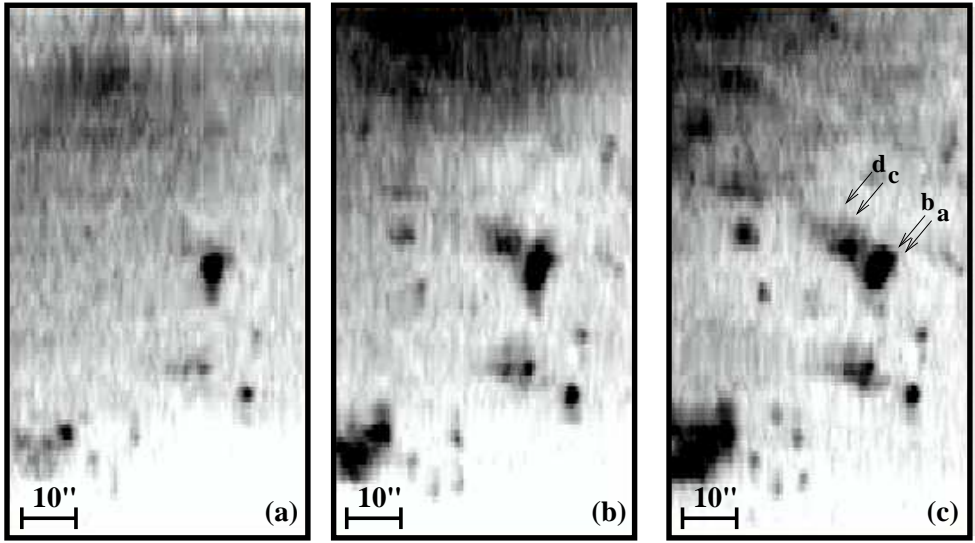}}  
\caption{Images in the light of the \nii\ lines in different ranges of
radial velocity a) \vhel\ = -31 to -27 \kms, b) -24 to -21 \kms\ and c)
-20 to -16 \kms\ are shown for Knot 32. The positions a-d
of the profiles in Fig. 3a--d are indicated in c).}
\label{fig4} 
\end{figure*}

\bsp  
  
\label{lastpage}  
  
\end{document}